# Dual-comb Spectroscopy of Ammonia Formation in Non-thermal Plasmas

Ibrahim Sadiek[1*], Adam J. Fleisher[2], Jakob Hayden[3], Xinyi Huang[3], Andreas Hugi[3], Richard Engeln[4], Norbert Lang[1], Jean-Pierre H. van Helden[1]

[1]Leibniz Institute for Plasma Science and Technology (INP), 17489 Greifswald, Germany
[2]Material Measurement Laboratory, National Institute of Standards and Technology, Gaithersburg, MD 20899, USA
[3]IRsweep AG, 8712 Staefa, Switzerland
[4]Plasma Physics Research, ASML Veldhoven, 5504 DR Veldhoven, The Netherlands

*To whom correspondence should be addressed; E-mail: ibrahim.sadiek@inp-greifswald.de

**Abstract:**

Plasma-activated chemical transformations promise the efficient synthesis of salient chemical products. However, the reaction pathways that lead to desirable products are often unknown, and key quantum-state-resolved information regarding the involved molecular species is lacking. Here we use quantum cascade laser dual-comb spectroscopy (QCL-DCS) to probe plasma-activated $NH_3$ generation with rotational and vibrational state resolution, quantifying state-specific number densities via broadband spectral analysis. The measurements reveal unique translational, rotational and vibrational temperatures for $NH_3$ products, indicative of a highly reactive, non-thermal environment. Ultimately, we postulate on the energy transfer mechanisms that explain trends in temperatures and number densities observed for $NH_3$ generated in low-pressure nitrogen-hydrogen ($N_2$–$H_2$) plasmas.

**Introduction**

Through its use in industrial fertilisers[1], ammonia ($NH_3$) makes an indispensable contribution to global agriculture by helping to feed about 50 % of the world's population[2]. Because of its critical place within our food supply chain, current research into plasma-activated $NH_3$ formation aims to reduce costs, improve distribution and increase efficiency as compared to the energy-intensive and ubiquitous Haber–Bosch process[3]. Ideally, plasma activated $NH_3$ production would be achieved by nitrogen ($N_2$) fixation from air, preferably at lower temperatures and pressures than are currently required. However, despite significant efforts, the energy yields reported for plasma-activated $NH_3$ formation remain more than one order-of-magnitude lower than that of the conventional Haber–Bosch process[4]. To overcome this deficit, it is crucial to improve our understanding of formation processes at the molecular level. Beyond its application to addressing the challenges of a global food supply chain, improved knowledge of plasma-activated $NH_3$ formation mechanisms could also impact our future energy needs by improving mitigation strategies applied to gas reprocessing in tokamak fusion reactors[5].

Because plasmas comprise high-energy electrons, ions, radicals and neutral molecules together in a confined environment, there exist myriad pathways that may lead to the breaking and formation of chemical bonds, and therefore a large number of possible chemical transformations. Generally, $NH_3$ formation in plasma reactors is ascribed to the stepwise hydrogenation of adsorbed nitrogen atoms (N), imidogen radicals (NH) and amino radicals ($NH_2$) found at various reactor surfaces[6,7]. If correct, this mechanism should yield $NH_3$ molecules in non-thermal equilibrium, an environment characterized by different temperatures being ascribed to different degrees of freedom of the molecule (translational, rotational and vibrational). Consequently, molecules in non-thermal equilibrium can exhibit enhanced or reduced rate constants for state-specific, vibrationally mediated reactions[8]. Therefore, to better understand the pathways to plasma-activated $NH_3$ formation, we require quantum-state-resolved information on the $NH_3$ molecule *in operando*. Although we focus on $NH_3$ in this work, quantum-state-resolved information and its impact on reaction pathways is of high importance to other plasma driven chemical syntheses[9]. This information is essential to develop and optimise new catalytic materials, which no longer have to enhance the dissociation of $N_2$, now taken care of by the plasma, but have to be tailored for maximum $NH_3$ production.

To date, several active laser diagnostics have been used to probe $NH_3$ molecules in plasma reactors. They are cavity-enhanced absorption spectroscopy[7,10], tuneable diode laser absorption spectroscopy[7,11] and quantum-cascade laser absorption spectroscopy[11–13]. These techniques rely upon stepwise scanning or sweeping the wavelength of a continuous-wave (CW) laser to measure the molecular absorption. Only a few absorption features are generally accessible to any one CW laser, limiting the scope of their application to only a sub-set of the complex plasma processes that occur between species within a non-thermal environment. To increase the spectral coverage and hence the number of quantum states that can be probed, external cavity diode lasers have been used. External cavity diode lasers, however, require *in situ* wavelength calibration to achieve the necessary spectral accuracy[14,15], and are often plagued by mode-hops. The overall measurement time to scan the single wavelength over a large range in fine increments presents another experimental challenge.

More broadly, in the mid-infrared wavelength region, direct frequency-comb Fourier transform spectroscopy has been used to detect multi-species in the effluent of a dielectric barrier discharge[16] and dual-comb spectroscopy has studied time-resolved spectra from a $CH_4$/He electric discharge[17]. In the visible wavelength region, dual-comb spectroscopy has also been used to detect trace amounts of the atomic species Rb and K after ejection from a laser-induced breakdown of a solid target[18]. But no previous demonstrations of comb spectroscopy have revealed a quantum-state-resolved picture of non-thermal plasmas applied to grand societal challenges like $N_2$ fixation to $NH_3$.

Here, we apply quantum cascade laser dual-comb spectroscopy (QCL-DCS)[19] at a wavelength near $\lambda$ = 9.4 µm to study $NH_3$ formation in a low-pressure, nitrogen-hydrogen ($N_2$–$H_2$) plasma. This study highlights several advantages of QCL-DCS as a diagnostic tool for probing complex environments like a non-thermal plasma, including high spectral resolution (4.5×10$^{-4}$ cm$^{-1}$ or 14 MHz) and broad spectral coverage (50 cm$^{-1}$). Previously, QCL-DCS has been applied to jet-cooled molecular expansions[20] and thermally populated gas samples at ambient temperature[21] to demonstrate high-resolution, and to condensed-phase biological systems to demonstrate fast (µs) acquisition rates[22]. Here we combine these two inherent advantages to rapidly perform high-resolution molecular spectroscopy of a complex, non-thermal environment inside of a research-grade industrial plasma reactor.

**Results**

**Plasma-activated $NH_3$ generation probed by dual-comb spectroscopy.**

We probe plasma-activated $NH_3$ generation via line-of-sight laser absorption spectroscopy in the long-wave infrared, near $\tilde{\nu}$ = 1060 cm$^{-1}$ ($\tilde{\nu} = c/\lambda$ in units of cm$^{-1}$ and $c$ is the speed of light). A schematic of the experimental set-up is shown in Fig. 1a. Briefly, the output from the first QCL comb with a repetition rate, $f_{\text{rep},1}$, of 7.417 GHz propagates along two distinct free-space beam paths, creating a probe beam and a reference beam. The probe beam is coupled to a multi-pass cell (path length of $L$ = 3.16 m) attached to the plasma reactor, and the reference beam bypasses the reactor. A second local-oscillator QCL comb with slightly different repetition rate ($f_{\text{rep},2} = f_{\text{rep},1} + \Delta f_{\text{rep}}$, where $\Delta f_{\text{rep}}$ = 2.1 MHz) is spatially overlapped with the probe and reference beams, respectively, at two different photodetectors, thus creating both probe and reference down-converted DCS signals (interferograms) for spectral analysis and optical power normalization. Additional details are provided in the Methods section.

The result is a transmission spectrum of $NH_3$, sampled at the frequencies of the first QCL comb. To perform high-resolution spectroscopy, both QCL combs are scanned by increasing the laser currents using a "step-sweep" approach[23], creating a set of 600 interleaved and normalized transmission spectra of $NH_3$ which together yield a composite transmission spectrum with an ultimate resolution of 4.5×10$^{-4}$ cm$^{-1}$ (14 MHz) achievable in seven minutes of total acquisition time.

The spectral region near $\tilde{\nu} \approx$ 1060 cm$^{-1}$ includes transitions from three different vibrational bands of $NH_3$. Using a labelling scheme which indicates the number of quanta excited in each of the $\nu_1\nu_2\nu_3\nu_4$ normal vibrational modes of $NH_3$, the transitions are described as (i) the fundamental $\nu_2 \leftarrow \nu_0$ band, corresponding to the 0100 ← 0000 transition, (ii) the $2\nu_2 \leftarrow \nu_2$ hot band, corresponding to the 0200 ← 0100 transition and (iii) the $\nu_2+\nu_4 \leftarrow \nu_4$ hot band, corresponding to the 0101 ← 0001 transition. Figure 1b shows a schematic of the vibrational levels probed here, with transition energies[24] listed in wavenumbers. Shown are the energy values of both the symmetric (grey) and antisymmetric (black) states with respect to molecular inversion. In the case of the symmetric states, the $\nu_4$ and $2\nu_2$ levels interchange their positions within the energy level diagram, as do the $\nu_2+\nu_4$ and $3\nu_2$ levels. Overall, the accessible vibrational states represent the first steps on a molecular vibrational ladder — or a series of energy levels which is useful for the study of overpopulation in non-thermal environments.

Here we generate the plasma in a research-grade, industrial reactor made of stainless steel by direct current (DC) discharge at a total power of 355 W ± 50 W. The discharge was located on a metal mesh of stainless steel in the top of the reactor. A workload made of stainless steel was also used to increase the discharge power and the stability of the plasma, and it was negatively biased relative to the reactor wall. A schematic of the plasma reactor is shown in Fig. 1a. Different mixtures of $N_2$ and $H_2$ gas precursors are delivered to the reactor at a constant mass flow rate of (500 ± 5) standard cubic centimeters per minute (sccm). The pressure inside the reactor is maintained at a constant value of 100 Pa ± 1 Pa using a back pressure controller and vacuum pump. The outer wall of the reactor is cooled to near room temperature at 295 K ± 1 K by a recirculating flow of water at 293.0 K ± 0.1 K, and the temperature of the inner wall ($T_{\text{wall}}$) is recorded by two temperature probes. Additionally, the temperature within the plasma at a blank stainless-steel working load ($T_{\text{load}}$) is also recorded using a third temperature probe.

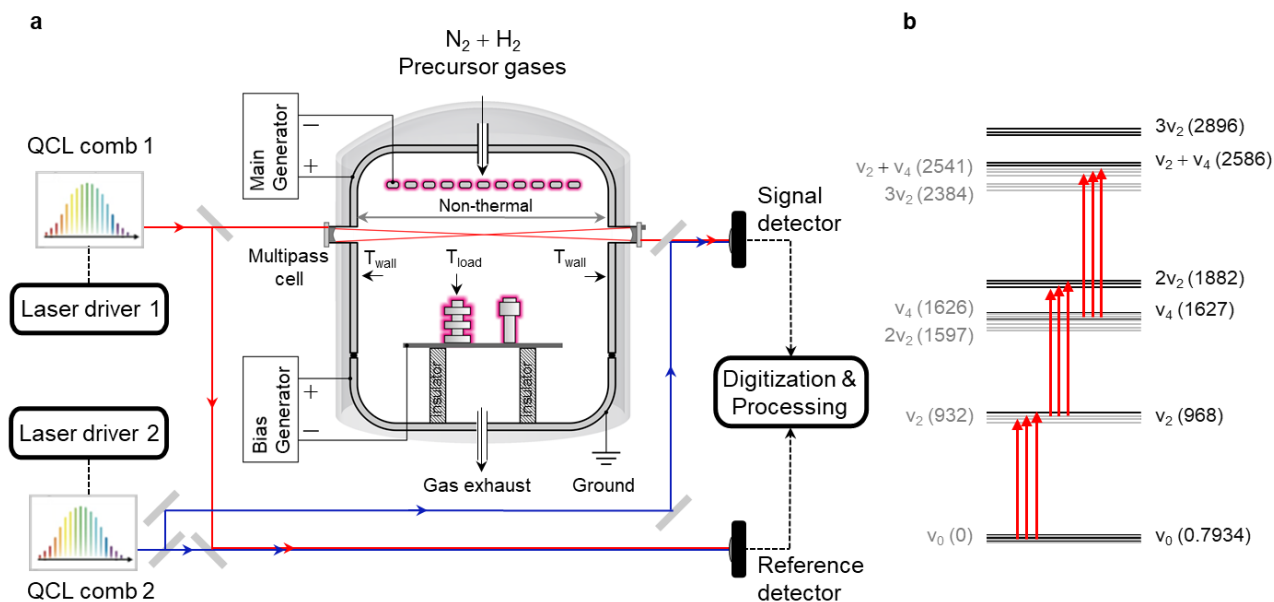

**Fig. 1 | Quantum cascade laser dual-comb spectroscopy (QCL-DCS). a** Schematic of the experimental setup. The probe beam from QCL comb 1 is coupled to a multi-pass cell attached to the plasma reactor, whereas the reference beam bypasses the reactor. Probe and reference combs are combined with copies of the local-oscillator QCL comb 2 at two different photodetectors. Dashed lines indicate electronic connections. Reactor temperatures $T_{\text{wall}}$ and $T_{\text{load}}$ were measured where indicated by the black arrows, and the non-thermal region (distance between the inner walls) of the reactor is indicated by the left-right arrow. **b** Energy level diagram for $NH_3$ at the $\nu_2$ vibration, as probed by QCL-DCS (vertical arrows). The transition energies (in wavenumber) of both the antisymmetric (black, horizontal) and symmetric (grey, horizontal) states are shown.

Figure 2a shows the broadband transmission spectrum of $NH_3$, measured for a plasma generated with precursor mass flow rates of 200 sccm of $H_2$ and 300 sccm of $N_2$. The strong absorption lines of the $NH_3$ $\nu_2 \leftarrow \nu_0$ fundamental band are saturated at mixing ratios of $H_2/N_2 \approx 1$ where the highest $NH_3$ yield is observed. The fitted spectral model, calculated using the HITRAN2020 database[25] and Voigt line shape functions, reveals absorption from the three different vibrational bands of $NH_3$ illustrated in Fig. 1b. We determine an $NH_3$ number density in the non-thermal region (the distance between the inner walls of the reactor), $n_{\text{n-th}}$, by fitting a spectroscopic model that accounts for possible non-thermal distributions by floating the translational temperature, $T_{\text{trans}}$, as well as the rotational temperature, $T_{\text{rot}}$, and vibrational temperature, $T_{\text{vib}}$, for the different vibrational bands. More details regarding the spectral model and fit are available in the Methods section. For the spectrum plotted in Fig. 2a, the resulting fit parameters and their estimated combined and relative uncertainties are listed in Table 1.

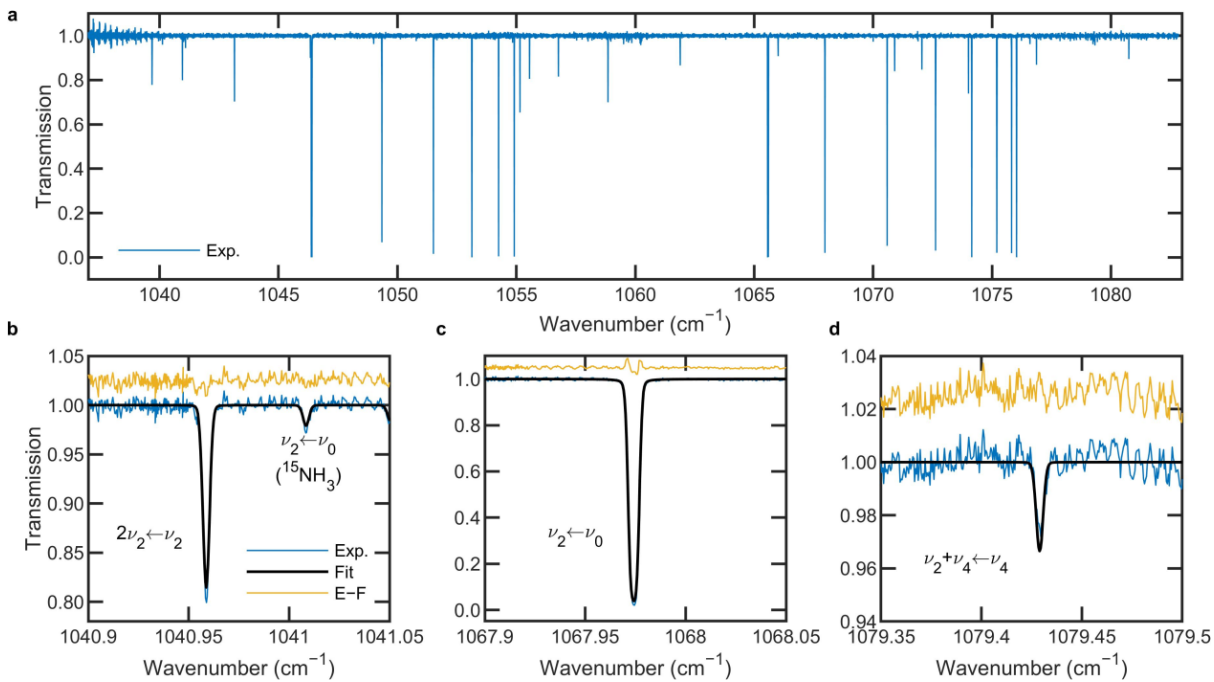

**Fig. 2 | QCL-DCS of plasma-activated NH$_3$ formation. a** Broadband transmission spectrum (Exp.; blue) of NH$_3$ generated in an N$_2$—H$_2$ plasma (H$_2$ mass flow: 200 sccm; N$_2$ mass flow: 300 sccm; pressure: 100 Pa ± 1 Pa). **b–d** Specific spectral regions showing transitions involving each $^{14}$NH$_3$ vibrational band, shown along with the fitted spectral model (Fit; black). Also shown are the Exp.-minus-Fit residuals (E–F; yellow), offset for clarity. **b** The $2\nu_2 \leftarrow \nu_2$ hot band, and the $\nu_2 \leftarrow \nu_0$ fundamental band observed for $^{15}$NH$_3$ at natural isotopic abundance. **c** The $\nu_2 \leftarrow \nu_0$ fundamental band. **d** The $\nu_2+\nu_4 \leftarrow \nu_4$ hot band.

**Table 1.** Fitted values for temperature ($T$) and non-thermal number density ($n_{\text{n-th}}$) parameters, derived from fitting the NH$_3$ transmission spectrum plotted in Fig. 2. Combined standard uncertainty and relative uncertainty are reported at the 1σ confidence level.

| Parameter | Fitted Value | Combined Uncertainty | Units | Relative Uncertainty | Units |
|---|---|---|---|---|---|
| $T_{\text{th}}$ | 310 | 20 | K | 6 | % |
| $T_{\text{trans}}$ | 456 | 10 | K | 2 | % |
| $T_{\text{rot}}$ ($\nu_0$) | 390 | 40 | K | 10 | % |
| $T_{\text{rot}}$ ($\nu_2$) | 460 | 30 | K | 7 | % |
| $T_{\text{rot}}$ ($\nu_4$) | 530 | 40 | K | 8 | % |
| $T_{\text{vib}}$ ($\nu_2$) | 419 | 13 | K | 3 | % |
| $T_{\text{vib}}$ ($\nu_4$) | 454 | 10 | K | 2 | % |
| $n_{\text{n-th}}$ | 3.6×10$^{14}$ | 2×10$^{13}$ | cm$^{-3}$ | 6 | % |

The transitions shown in Fig. 2b–d also serve to highlight an advantage of our broadband approach. As an example, if we assume an average global temperature for the non-thermal region of $T_{\text{trans}} = T_{\text{rot}} = T_{\text{vib}}$ = 450 K, a line-by-line analysis of these respective transitions would yield significantly different NH$_3$ number densities, $n_{\text{n-th}}$, when fitted for each respective spectral region. In such a scenario, single-transition fits of $n_{\text{n-th}}$ would differ between the spectral regions shown in Fig. 2b–d by a factor of six — potentially biasing conclusions on the efficiency of NH$_3$ formation in non-thermal plasmas and impeding a quantitative comparison between different reactors.

**NH$_3$ number densities measured for different H$_2$ mass flow fractions.**

In Figure 3 we show the number density of NH$_3$ in the non-thermal region ($n_\text{n-th}$), obtained from our broadband spectral fits and plotted as a function of the H$_2$ mass flow fraction. The H$_2$ mass flow fraction is defined as $\phi_{H_2} = \dot{m}_{H_2}/(\dot{m}_{H_2} + \dot{m}_{N_2})$, where $\dot{m}$ is the mass flow of each precursor gas. The figure comprises data retrieved from the fitting of 27 unique spectra measured at different values of $\phi_{H_2}$. The spectra were collected in two series, proceeding from either low to high values of $\phi_{H_2}$ (left-to-right, blue circles) or high to low values of $\phi_{H_2}$ (right-to-left, red triangles). The highest NH$_3$ yield is observed on the H$_2$-deficient side of the data set agrees with measurements performed on similar plasma reactors[12]. However, the maximum NH$_3$ yield has also been observed in other types of reactors to be on the opposite, H$_2$-abundant side[6,7,10,11]. The position of the maximum and the observed asymmetry with respect to $\phi_{H_2}$ is influenced by the reactor wall material and process pressure, as well as the plasma electron energy and density[7,26,27]. The reported NH$_3$ number densities span more than two orders of magnitude (see insets in Fig. 3), illustrating the wide dynamic range capabilities of QCL-DCS as a laser-based plasma diagnostic technique.

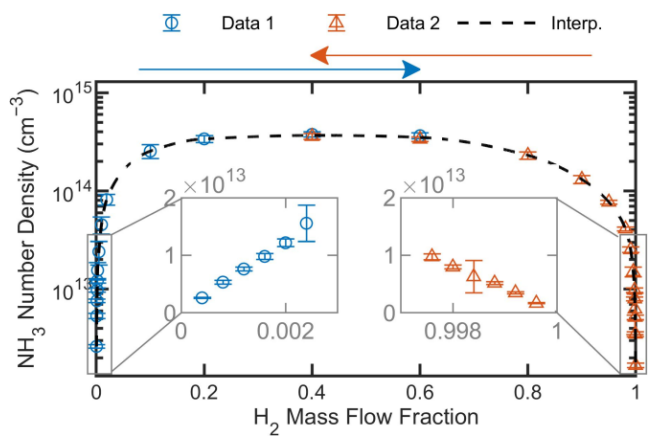

**Fig. 3 | NH$_3$ number density in the non-thermal region, $n_\text{n-th}$, as a function of the H$_2$ mass flow fraction, $\phi_{H_2}$.** Mixed precursor gases were maintained inside the plasma reactor at a fixed pressure of 100 Pa ± 1 Pa. Measurements for the two data series (Data 1 and Data 2; blue circles and red triangles) proceeded in two distinct $\phi_{H_2}$ directions (blue and red arrows). A modified Akima interpolation model[25] (Interp.; black dashed line) is shown to illustrate the asymmetry in the observed data set, and zoomed-in traces near $\phi_{H_2}$ = 0 and $\phi_{H_2}$ = 1 are plotted as grey insets (linear y-axis). Error bars are combined standard uncertainty.

**Non-thermal population of NH$_3$ states.**

In Figure 4 we show the fitted temperatures that partition population amongst the translational, rotational and vibrational states of NH$_3$ found within the non-thermal region of the plasma reactor. Plotted in Fig. 4a are the fitted values of $T_\text{vib}$ vs. $\phi_{H_2}$. Generally, values of $T_\text{vib}$ are between 400 K and 500 K, and we observe $T_\text{vib}^{\nu_4} > T_\text{vib}^{\nu_2}$. Also plotted is a smoothing spline fitted to the average measured value of $T_\text{wall}$. Figure 4b shows the fitted values of $T_\text{rot}$ plotted vs. $\phi_{H_2}$. The observed rotational temperatures reveal a general trend, where $T_{\text{rot}}^{\nu_4} > T_{\text{rot}}^{\nu_2} > T_{\text{rot}}^{\nu_0}$. Two smoothing splines appear in Fig. 4b, each fitted to the average values of $T_{\text{load}}$ and $T_{\text{wall}}$, respectively. No external heating was applied to the plasma reactor, and $T_{\text{load}} > T_{\text{wall}}$ is due to the negative bias applied to the working load. For the $2\nu_2 \leftarrow \nu_2$ and $\nu_2+\nu_4 \leftarrow \nu_4$ hot bands, values of $T_{\text{vib}}$ and $T_{\text{rot}}$ are only reported for spectra with an observed signal-to-noise ratio (SNR) greater than or equal to four for the strongest rovibrational transition within each respective hot band. The choice of SNR > 4 is rather arbitrary, but roughly corresponds with the minimum SNR required for the fitting of band temperatures to converge.

Plotted in Fig. 4c are the fitted values for NH₃ translational temperatures in the non-thermal ($T_{\text{trans}}$) and thermal ($T_{\text{th}}$, in regions beyond the inner walls of the reactor) regions of the line-of-sight. In the thermal region, we assume a single temperature for all degrees of freedom (i.e., $T_{\text{th}} = T_{\text{rot}} = T_{\text{vib}}$), and the fitted results reveal a temperature for the thermal region that is consistent with the room temperature and the outer plasma-reactor walls, $T_{\text{th}} \approx 300$ K. The non-thermal region, where NH₃ is formed, shows a similar trend in $T_{\text{trans}}$ vs. $\phi_{H_2}$ to that observed for the NH₃ number density, $n_{\text{n-th}}$, in Fig. 3.

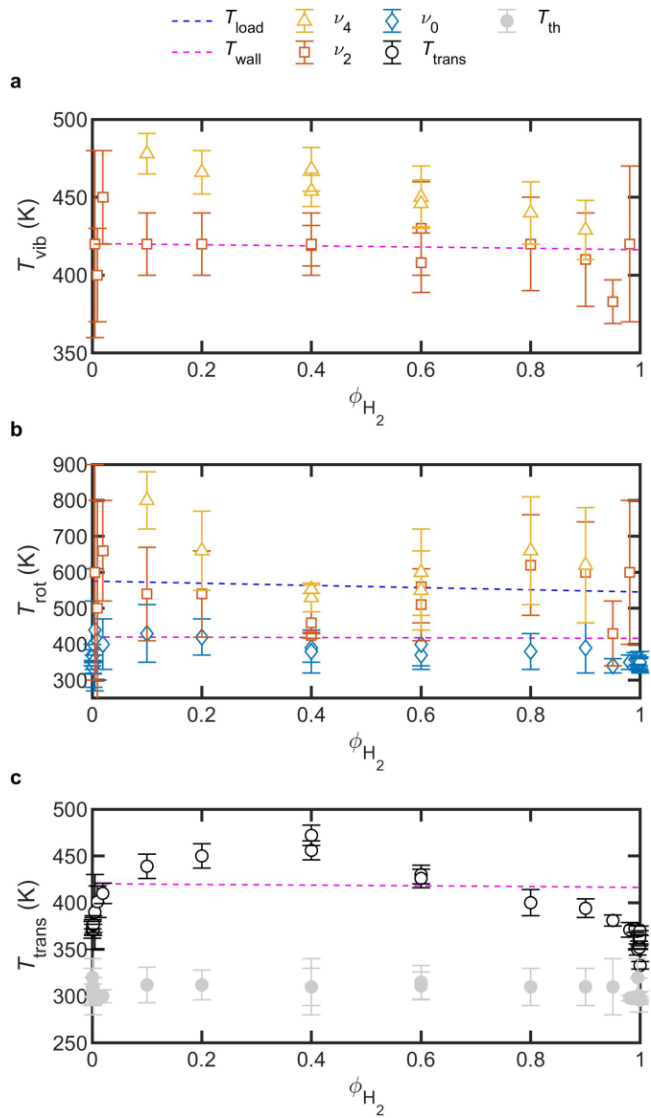

**Fig. 4 | Distinct temperatures for plasma-activated NH₃ generation. a** Vibrational temperatures ($T_{\text{vib}}$) for the $\nu_4$ (yellow triangles) and $\nu_2$ (red squares) progressions, plotted vs. $\phi_{H_2}$. **b** Rotational temperatures ($T_{\text{rot}}$) for the transitions beginning in $\nu_4$ (yellow triangles), $\nu_2$ (red squares) and $\nu_0$ (blue diamonds) states, plotted vs. $\phi_{H_2}$. **c** Translational temperatures in the non-thermal ($T_{\text{trans}}$, black circles) and thermal ($T_{\text{th}}$, gray filled circles) regions, plotted vs. $\phi_{H_2}$. Also plotted in **a**–**c** are the smoothed values of $T_{\text{load}}$ (blue dashed line) and $T_{\text{wall}}$ (magenta dashed lines). The smoothed values of $T_{\text{load}}$ are only visible in **b**, where higher temperatures are shown.

## Discussion

The N₂–H₂ plasmas generated here are a case-study of plasma-activated N₂ fixation to NH₃. Therefore, it is instructive to postulate on the energy transfer dynamics that yield the non-thermal populations of rotational and vibrational states observed for NH₃ products.

We partially attribute the measured non-thermal population of excited states — observed in the gas phase — to surface association reactions of radicals and atoms adsorbed on the reactor walls. These surface association reactions then release NH₃ into the gas phase in vibrationally excited states. In addition to direct NH₃ formation, surface association reactions can also form rovibrationally excited H₂, as has already been reported in expanding plasmas[29], and electron impact reactions can yield other vibrationally excited molecules, such as NH₃[30].

Regardless of the formation mechanism, vibrationally excited NH₃ in the gas phase is not likely to undergo rapid population redistribution through a vibrational-vibrational (V-V) energy transfer mechanism. Instead, and somewhat unique to NH₃, experimental evidence suggests that vibrational-translational (V-T), vibrational-rotational (V-R), rotational-rotational (R-R) and rotational-translational (R-T) relaxation mechanisms are more likely[31–36]. This picture is consistent with the relatively low vibrational temperatures observed for NH₃ in Fig. 4a, where we find $T_{\text{vib}} < T_{\text{rot}}$ for both the $\nu_4$ and $\nu_2$ progressions. Indeed, NH₃ behaves differently than other molecules like CO₂ which exhibit rapid V-V relaxation processes along a vibrational ladder, leading to higher vibrational temperatures observed for CO₂ formed in non-thermal plasmas[37] than are observed here for NH₃.

In the following discussion, we focus on two more key observations: (i) The rotational temperature of the $\nu_0$ state is lower than the translational temperature, and (ii) The translational temperature appears to follow the same trend vs. $\phi_{H_2}$ as the non-thermal NH₃ number density.

Together, these observations further point to the role of rapid V-T, V-R, R-R and R-T energy transfer mechanisms in dictating the non-thermal population distribution of NH₃ products. Initially, vibrationally excited NH₃ is rapidly deactivated to translational and rotational degrees of freedom. Following V-V relaxation which takes place on longer time scales, any remaining vibrational energy in the lowest vibrational level of a given mode — here, $\nu_2$ or $\nu_4$ — is further dissipated into translational energy via V-T relaxation. If the translational degree of freedom is taken to be the main sink for excess vibrational energy contained in the initial NH₃ products, this qualitatively explains observation (i), where $T_{\text{rot}}^{\nu_0} < T_{\text{trans}}$.

Furthermore, NH₃ is proposed as the major sink for excess vibrational energy found in all the plasma-activated molecules, due to relaxation via near-resonant V-V energy transfer between NH₃ and a vibrationally excited N₂ or H₂ collider. Once the excess energy is transferred to NH₃, the above-mentioned mechanisms of V-T, V-R, R-R and R-T relaxation dominate. In such a scenario, energy is most efficiently dissipated into translation motion through NH₃ relaxation channels. Ammonia is a polar molecule with a steep intermolecular potential energy curve for collisional processes, leading to efficient relaxation of internal energy into heat[31,32]. This suggests that differences in buffer gas composition would affect efficiency, and hence chemical reactivity within $N_2$—$H_2$ plasmas. Indeed, we anticipate that vibrationally excited $N_2$ ($\tilde{v}$ = 2311 cm$^{-1}$) and $H_2$ ($\tilde{v}$ = 4160 cm$^{-1}$) are both formed in the plasma, and could transfer their energy via rapid V-V mechanisms into $NH_3$ rather than dissipating it by self-collision and internal V-T processes. The number of collisions required to achieve self-deactivation via V-T for $NH_3$, $H_2$ and $N_2$, respectively, is approximately 5, 10$^7$ and 10$^9$ at a temperature of 300 K[33]. For comparison, the deactivation of $NH_3$ by collision with $N_2$ is measured to require only 670 collisions[32]. From these values, we would expect that deactivation of $NH_3$ by collision with $H_2$ would take greater than 670 collisions, as could also be inferred from looking at the respective vibrational frequencies of $H_2$ and $N_2$ (which differ by almost a factor of two). Therefore, the most efficient route to deactivation of vibrationally excited molecules in the plasma appears to be through collisions with $NH_3$. This hypothesis is consistent with the following observation (ii), that $T_{\text{trans}}$ appears to follow an asymmetry trend vs. $\phi_{H_2}$ that is qualitatively like that of $NH_3$ number density. As the energy sink molecule in our $N_2$–$H_2$ plasmas, greater number densities of $NH_3$ result in a higher $T_{\text{trans}}$.

The observation of similar trends in $T_{\text{trans}}$ and $n_{\text{n-th}}$ when plotted vs. $\phi_{H_2}$ provides additional support for the argument that V-T processes are largely responsible for depleting the anticipated initial population of excited vibrational states that occurs immediately following $NH_3$ formation at the reactor surfaces. Contrast this trend with the impact that collisions between plasma-generated vibrationally excited $NH_3$ and electrons would have on the observed steady-state populations: If processes such as scattering, momentum transfer, dissociation, and excitation of rotational and vibrational degrees of freedom were induced by electron-$NH_3$ collisions, we would expect a monotonic change in $T_{\text{trans}}$ vs. $\phi_{H_2}$.[30] Additionally, the rate coefficient for the dissociation of $NH_3$ by electrons — calculated at an electron temperature of 0.31 eV observed for similar plasma reactors[38] — is roughly three orders of magnitude lower than the predicted V-T relaxation rate[26]. Thus, V-T is likely the dominant relaxation process for $NH_3$ once generated in $N_2$–$H_2$ plasma.

**Outlook**

We demonstrate that quantum cascade laser dual-comb spectroscopy (QCL-DCS) can provide precision measurements of number densities and non-thermal population distributions across the translational, rotational and vibrational degrees of freedom of molecules confined to a reactive plasma environment. This is achieved here for a broad optical bandwidth by a combination of rapid spectra acquisition and high spectral resolution. With other chip-based laser sources like the long-wave infrared QCLs used here, DCS could also be employed as a plasma diagnostic in the mid-wave infrared using interband cascade lasers[39] or in the THz regime using QCLs[40]. When combined with injection-locked electro-optic comb generators in the short-wave infrared[41], it may become possible to use DCS as a plasma diagnostic anywhere across the wavelength range of 1 µm to 100 µm, choosing between a series of distinct yet compact instruments. This would advance the field of laser-based plasma diagnostics beyond the current state-of-the-art — where only a few pre-selected rovibrational transitions of a single molecular species are observed — by enabling observations of several vibrational bands of multiple species, without sacrificing spectral resolution or measurement speed.

In this demonstration, we begin to unravel the complex energy transfer mechanisms that result from a non-thermal population of energy levels in plasma-activated ammonia. Specifically, we highlight the role of various energy transfer mechanisms in partitioning population densities amongst quantized states, a process which significantly affects chemical reactivity. This opens the door to further systematic studies of plasma-activated processes involving $NH_3$ generation, water-enhanced $NH_3$ synthesis[42], or the conversion of carbon dioxide to high value-added chemical products like renewable fuels. Furthermore, such a quantum-state-resolved picture of molecules within reactive environments will be of high importance for understanding other plasma driven chemical syntheses and transformations.

**Methods**

**Quantum cascade laser dual-comb spectroscopy setup.**

The dual-comb source (IRsweep IRis-core) emitted in the spectral range from 1035 cm$^{-1}$ to 1085 cm$^{-1}$ with repetition rates of $f_{rep} \approx 7.417$ GHz, a difference in repetition rates of $\Delta f_{rep}$ = 2.1 MHz and average output optical powers ≥100 mW. The QCL outputs were attenuated by approximately tenfold using neutral density filters and aligned to create two dual-comb paths: reference and probe. Polarizers were used in each recombined QCL beam path to match interferogram intensities, and the dual-comb beams were focused onto photodetectors using off-axis parabolic mirrors with focal lengths of 25.4 mm.

Transmission spectra, $T(\tilde{v})$ were calculated by squaring the ratio $I^b(\tilde{v})/I_0^b(\tilde{v})$, where $I^b(\tilde{v})$ is the intensity of the multi-heterodyne beat notes measured after passing through a plasma containing both $N_2$ and $H_2$ (the sample spectrum) and $I_0^b(\tilde{v})$ is the intensity measured in a pure $N_2$ or $H_2$ plasma (the background spectrum) where no $NH_3$ is formed. The squaring of the ratio is required when only one of the two interfering combs in the sample channel passes through the sample, thus creating a phase-sensitive configuration[43]. To suppress laser intensity noise and frequency noise, all measured intensities are normalized by the simultaneously measured intensities in the reference channel where both combs bypass the reactor.

High-resolution spectra were obtained by spectral interleaving of 600 measurements following the "step-sweep" approach[23], yielding a spectral point spacing of 4.6×10$^{-4}$ cm$^{-1}$ (14 MHz) in a total measurement time of seven minutes. The absolute frequency axis was calibrated by matching the offset frequency, $f_{off}$ and $f_{rep}$ of the first measurement step to a pair of $NH_3$ line positions retrieved from the HITRAN2020 database[25]. To correct for a residual drift of the measured wavenumber axis with interleaving step number, the measured changes in $f_{off}$ in every step were corrected by a constant factor of 0.9942 and 0.9972, respectively, for measurements taken with two slightly different stabilization times on different days. These factors were determined as the linear slope in the difference between measured and reference absorption peak positions of $NH_3$ from the HITRAN2020 database[25], plotted against the step number at which the absorption peak was measured. Hence, three fitting parameters ($f_{rep}$, $f_{off}$, and $\Delta f_{off}$) were used to assign approximately 125 000 spectral datapoints. After such calibration, the difference between found peak positions and those listed in the HITRAN2020 database[25] were < 4.3×10$^{-4}$ cm$^{-1}$ (or 13 MHz).

The interleaved spectra of plasma samples showed a constant offset (>5 % transmission) in amplitude and linear slope in phase. These background signals varied on timescales from minutes to hours, as well as on a timescale of seconds, or between interleaving steps. The slower variations are likely due to thermal drifts. Since the spectrometer and the reactor are not mechanically connected, thermal expansion in the plasma reactor can change detector alignment and therefore signal levels. The origin of the fast drifts could not be identified. The constant offset in amplitude and linear slope in phase were both removed in post-processing from every interleaving step. To this end, absorption features were first masked out. Then, the median value of transmission of the remaining spectral datapoints was subtracted, as well as a linear function fitted to the phase. This procedure reduced the root-mean-square noise on the measured transmission to 0.0045, with the remaining noise dominated by optical fringing in the multi-pass cell, which was further fitted by a series of polynomial baseline functions.

**Broadband spectral model and analysis.**

In absorption spectroscopy, the laser intensity spectrum, $I(\tilde{v})$, after having propagated along an absorption pathlength, $L$, and being normalized by a background intensity spectrum, $I_0(\tilde{v})$, follows an exponential decay described by the Beer–Lambert law. Here, the ratio $I(\tilde{v})/I_0(\tilde{v})$ is taken to be the experimental observable for asynchronous DCS, $\{I^b(\tilde{v})/I_0^b(\tilde{v})\}^2$. The Beer–Lambert law is stated in Eq. (1), and includes a total number density of absorbers, $n$, a spectral line intensity, $S_{ij}$, for a transition connecting a lower state, $i$, with an upper state, $j$, and an area-normalized line shape function, $g(\tilde{v})$:

$$\int \ln\left(\frac{I_0(\tilde{v})}{I(\tilde{v})}\right) d\tilde{v} = nS_{ij}L \int g(\tilde{v})d\tilde{v} \tag{1}$$

Above, $\tilde{v}$ is frequency in wavenumbers. The transition intensity, $S_{ij}$, is related to the difference in number density between the lower and upper states[44] by the following equation:

$$S_{ij} = I_a \frac{1}{n}(n_i B_{ij} - n_j B_{ji})\frac{h\tilde{v}_{ij}}{c} \tag{2}$$

In Eq. (2), $I_a$ is the isotopic abundance of the species involved in the specific transition, $n_i$ is the lower-state number density, $n_j$ is the upper-state number density, $B_{ij}$ and $B_{ji}$ are the Einstein $B$-coefficients for induced absorption and emission, respectively, $h$ is the Plank constant, $\tilde{v}_{ij}$ is the transition frequency, and $c$ is the speed of light. Note that $g_i B_{ij} = g_j B_{ji}$ and $A_{ij} = 8\pi h\tilde{v}_{ij}^3 B_{ji}$, where $g_i$ is the lower-state statistical weight, $g_j$ is the upper-state statistical weight, and $A_{ij}$ is the Einstein $A$-coefficient for spontaneous emission.

Following the procedure of Klarenaar et al.[37], we write the number density $n$ for each energy level $l$, $n_l$, where $l = i$ or $l = j$, as:

$$n_i = n\phi_{\text{rot},J}\prod_m \phi_{\text{vib},v_m} \tag{3}$$

Above, $J$ is the rotational quantum number and $m$ is the index of vibrational modes ($v_m$ = 1, 2, 3, or 4). The fraction of molecules, $n_l/n$, in both rotational state $J$, $\phi_{\text{rot},J}$, and vibrational mode $v_m$, $\phi_{\text{vib},v_m}$, is normalized by the total internal partition sum, $Q_{\text{tot}} = Q_{\text{rot}}Q_{\text{vib}}$. Here, the internal partition sums for rotation and vibration are[44–46]:

$$Q_{\text{rot}}(T_{\text{rot}}) = \sum_i (2J+1)g_s g_{\text{in}} \exp\left(-\frac{hcE_J}{k_B T_{\text{rot}}}\right) \tag{4}$$

$$Q_{\text{vib}}(T_{\text{vib}}) = \prod_{v_m} \left(1 - \exp\left(-\frac{hcG_{v_m}}{k_B T_{\text{vib}}}\right)\right)^{-g_{v_m}} \quad (5)$$

In Eqs. (4)–(5), $T_{\text{rot}}$ is the rotational temperature, $T_{\text{vib}}$ is the vibrational temperature, $g_s$ and $g_{in}$ are the state-dependent and state-independent weights, $g_{v_m}$ is the degeneracy for the fundamental vibration $v_m$, and $G_{v_m}$ is the term symbol for vibration $v_m$. To calculate $Q_{\text{rot}}$ for ammonia (NH$_3$), we use values for $g_s$ and $g_{in}$ from Šimečková et al.[44] and sum over all rotational states listed in ExoMol[47,48], up to $J_{\text{max}}$ = 43. To calculate $Q_{\text{vib}}$ for NH$_3$, we use values for $G_{v_m}$ from Polyansky et al.[49] and degeneracy factors $g_{v_m}$ derived from the $D_{3h}$ point group[50]. At values of $T$ < 1000 K, when letting $T = T_{\text{rot}} = T_{\text{vib}}$, our calculation of $S_{ij}$ using Eq. (2) has a relative deviation from the temperature-dependent values calculated from HITRAN2020 parameters[25] of <6 %. This bias was corrected for each vibrational band and is ascribed to the use of an incomplete list of energy levels in our total partition function summations.

For the line shape function, $g(\tilde{v})$, we use a Voigt function with a Doppler-broadened half-width at half-maximum of:

$$\Gamma_D = \tilde{v} \sqrt{\frac{2 N_A k_B T_{\text{trans}} \ln(2)}{M c^2}} \quad (6)$$

where $N_A$ is the Avogadro constant, $T_{\text{trans}}$ is the translational temperature, and $M$ is the molecular molar mass. Here we assume a single, shared value for $T_{\text{trans}}$ across all transitions. The Lorentzian term for the Voigt function was calculated using the measured gas pressure and the air-broadening coefficients for NH$_3$ from HITRAN2020[22].

For the non-thermal region of the plasma reactor, we use the above equations to model and fit the absorption by 86 total transitions belonging to the $v_2 \leftarrow v_0$ fundamental band (20 transitions), the $2v_2 \leftarrow v_2$ hot band (15 transitions), and the $v_2+v_4 \leftarrow v_4$ hot band (51 transitions). All other transitions meeting an intensity threshold criterion of $(5 \times 10^{-5}) \times S_{ij,\text{max}}(T_{\text{rot}}, T_{vib})$, where $S_{ij,\text{max}}$ is the maximum temperature-dependent transition intensity generated from within the list of the 86 targeted lines, are simulated assuming $T = T_{\text{trans}} = T_{\text{rot}} = T_{\text{vib}}$ at a temperature equal to the fitted value of $T_{\text{trans}}$.

Based on machine drawings of the physical dimensions of the plasma reactor, we estimate the single-pass path length of the non-thermal region along the line-of-sight to be 64.0 cm — equivalent to the physical distance between the hot inner walls of the reactor. For a single-pass total optical path length of 79.0 cm ± 0.5 cm, we estimate a thermal region beyond the hot inner walls to be 15.0 cm in length, resulting in a fractional thermal path length of $f_{\text{th}}$ = 0.190. Again, following the procedure of Klarenaar et al.[37], we fit a single thermal temperature, $T_{\text{th}}$, for the thermal region with a lower bound equal to the observed room temperature of 295 K and an upper bound equal to the translational temperature of the non-thermal region, $T_{\text{trans}}$. The combined model for the observed transmission signal is then the product of the respective transmission models for the thermal and non-thermal regions. Thermal and non-thermal number densities are calculated assuming the ideal gas law, using either $T_{\text{th}}$ or $T_{\text{trans}}$.

**Uncertainty analysis.**

We adopt a probabilistic approach to the uncertainty propagation[51], using Monte Carlo simulation methods and fitting to produce a distribution of output values that are the result from models generated by randomly drawn inputs. Components of the spectral reference data used to model NH$_3$ absorption are evaluated at randomly selected values assuming a normal distribution with a standard deviation equal to the upper-limit HITRAN2020 uncertainty codes[25]. For example, the transition intensity error codes for most of the lines included in our model have a relative uncertainty of <20 %. For the physical input parameters optical path length and pressure, we also randomly draw values from normal distributions. A summary of model input parameters with sizeable standard deviations, listed as relative uncertainties, is shown in Table 2.

**Table 2.** Table of model input parameters and their relative uncertainties.

| Parameter | Relative Uncertainty (%) | Description | Source |
|---|---|---|---|
| $p$ | 0.2 | Sample pressure | Manufacturer specification |
| $L$ | 0.6 | Optical path length | Machine drawing |
| $S_{ij,\text{calc}}^{\nu_2}$ | < 0.6 | Calculated model from Eq. (2), fundamental band | Scatter relative to HITRAN2020[25] |
| $S_{ij,\text{calc}}^{2\nu_2}$ | < 1.1 | Calculated model from Eq. (2), $2\nu_2$ hot band | Scatter relative to HITRAN2020[25] |
| $S_{ij,\text{calc}}^{\nu_2+\nu_4}$ | < 5.6 | Calculated model from Eq. (2), $\nu_2+\nu_2$ hot band | Scatter relative to HITRAN2020[25] |
| $S_{ij,\text{HITRAN}}$ | 10.0 | Select transition intensities, fundamental band | Ref. 25; Ref. 52 |
| $S_{ij,\text{HITRAN}}$ | 20.0 | Most transition intensities, all bands | Ref. 25; Ref. 53 |

During fitting, we use the HITRAN2020 isotopic abundance value to model all $^{15}$NH$_3$ lines. In determining $T_{\text{trans}}$ from the Doppler-broadened line widths, uncertainties in $\tilde{\nu}_{ij}$ and $M$ are considered negligible. Also, at the sample pressure of 100 Pa ± 1 Pa, uncertainties in the collisional-broadening parameters are considered negligible. Finally, the uncertainties in the energies of the individual energy levels are considered negligible. At the observed experimental signal-to-noise ratio, we find no systematic deviations in the residuals that would indicate the need to simulate line profiles beyond the Voigt profile, and we maintain a fixed value for the fractional thermal path length, $f_{\text{th}}$, when drawing random values for the total path length, $L$.

For the spectra where lines from all three of the vibrational bands have a high signal-to-noise ratio, we model and fit 100 unique Monte Carlo model simulations drawn from the uncertain input parameters. The reported values and uncertainties for each floated parameter are taken to be the mean values and standard deviations resulting from the 100 fits. The Monte Carlo routine also included variations in the initial values for each floated parameter. For spectra where only one or two vibrational bands are of sufficiently high SNR to fit, the number of Monte Carlo simulations and fits is reduced to 10.

When the signal-to-noise ratio (SNR) becomes small for a specific vibrational band (i.e., when SNR < 4 for the strongest transitions), we fix the values for $T_{\text{rot}}$ and $T_{\text{vib}}$ for that band which are randomly drawn from the weighted mean and weighted standard deviation of the fitted values from the other spectra with SNR ≥ 4.

Tables of initial parameters and drawing distributions used for each fitted spectrum, along with the mean values and standard deviations that resulted from the Monte Carlo simulation and fitting routine, are provided as Supplementary Information.

**Preliminary line-by-line spectral analysis.**

Prior to the broadband spectral modelling and fitting described above and reported in the Results section, a preliminary line-by-line analysis was useful in estimating initial guesses for parameters like $T_{\text{trans}}$ and $T_{\text{rot}}$, and for identifying transitions — particularly hot-band transitions — that required manual adjustments to their HITRAN2020[25] frequencies to accommodate our assigned wavenumber axis. A list of transition frequency adjustments applied to our broadband model is provided as Supplementary Information.

Here we focus on a Boltzmann plot analysis to determine initial values of $T_{\text{rot}}$. By plotting the natural logarithm of the lower-state number density, $n_i$, normalized by the lower-state statistical weight, $g_i$ (i.e., the quantity $\ln(n_i/g_i)$), versus the lower-state energy, $E_i$, rotational temperatures could be estimated from the band-specific Boltzmann plot slopes. The Boltzmann plot analysis is presented here in Fig. 5.

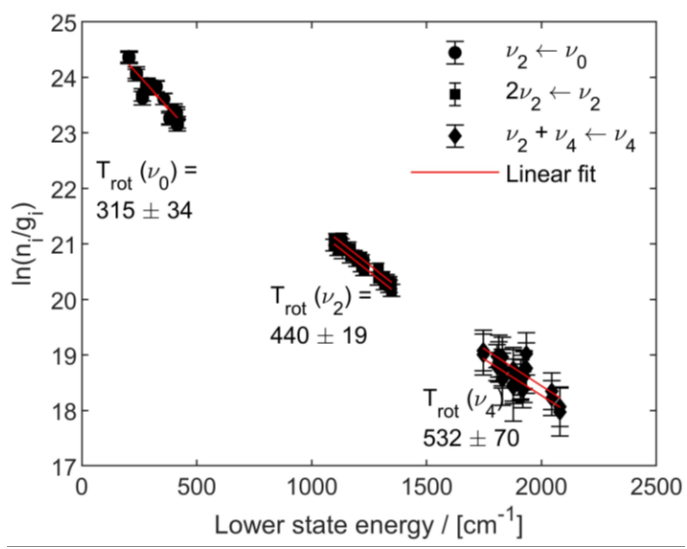

**Fig. 5 | Boltzmann plot to estimate band-specific rotational temperatures.** Rotational temperatures, $T_{\text{rot}}$, were estimated for three rovibrational bands of NH₃, including the $\nu_2 \leftarrow \nu_0$ fundamental band (black dots), the $2\nu_2 \leftarrow \nu_2$ hot band (black squares), and the $\nu_4+\nu_2 \leftarrow \nu_4$ hot band (black diamonds). Linear fits are also shown (red lines). Two data sets for each vibrational band are presented, each of which was collected at $\phi_{H_2}$ = 0.6 and recorded on successive days. Error bars represent 1σ fit precision in the area of the line profile (type-A evaluation only).

To create Fig. 5, individual lines were fit with a Voigt line shape function, with the fit parameters including the Doppler-broadened half-width at half-maximum and the line center. The translational temperature ($T_{\text{trans}}$) was evaluated from the Doppler width using Eq. (6), while $n_i$ was determined from the integral of the absorption profile and the HITRAN2020 parameters $A_{ij}$ and $\tilde{\nu}_{ij}$.

More detailed Boltzmann plot expressions are introduced below. The statistical-weight-normalized number density in state $i$ is described by the Boltzmann law:

$$\frac{n_i}{g_i} = \frac{n}{Q_{\text{tot}}(T)} \exp\left(-\frac{hcE_i}{k_B T}\right) \quad (7)$$

Remembering the expressions $g_i B_{ij} = g_j B_{ji}$ and $A_{ij} = 8\pi h \tilde{v}_{ij}^3 B_{ji}$ noted earlier, and by rearrangement of the Beer-Lambert law expressed in Eq. (1) and the transition intensity defined in Eq. (2)[44], we can also write the ratio $n_i/g_i$ in terms of the experimental observables $I_0(\tilde{v})$ and $I(\tilde{v})$:

$$\frac{n_i}{g_i} = \left(\frac{8\pi c \tilde{v}_{ij}^2}{g_j A_{ij} L}\right) \int \ln\left(\frac{I_0(\tilde{v})}{I(\tilde{v})}\right) d\tilde{v} \quad (8)$$

In Eq. (8), we assume that $\left(n_i \frac{g_j}{g_i} - n_j\right) \approx n_i \frac{g_j}{g_i}$ if $n_i \gg n_j$. This assumption is reasonable for the transitions studied here when the sample temperature is near 300 K. At higher temperatures, however, this is not a safe assumption, and therefore the rotational temperatures estimated from the Boltzmann plot in Fig. 5 are used as initial values in the broadband spectral model and fit. Taking the natural logarithm of both sides of Eq. (7) — and using Eq. (8) along with reference quantities from HITRAN2020[22] to calculate $n_i/g_i$ from the fits of our experimental data — yields the desired Eq. (9), where the slopes of the data plotted in Fig. 5 are inversely proportional to the estimated rotational temperatures for each vibrational band:

$$\ln\left(\frac{n_i}{g_i}\right) = -\frac{hc}{k_B T} E_i + \ln\left(\frac{n}{Q_{\text{tot}}(T)}\right) \quad (9)$$

Note that the initial Boltzmann plot analysis does not account for the thermal population outside the reactor core.

**Acknowledgement.** The INP acknowledges funding from the Deutsche Forschungsgemeinschaft (DFG, German Research Foundation) - project No SA 4483/1–1. IRsweep acknowledges funding from the European Union's Horizon 2020 research and innovation programme under the Marie Skłodowska-Curie grant agreement No 101032761. AJF acknowledges both NIST and the INP for supporting a research visit to INP. We thank J. T. Hodges, E. A. Adkins, B. R. Washburn, M. Becker, and A. Foltynowicz for commenting on the manuscript.

**Author contributions.**

According to CRediT (Contributor Roles Taxonomy), **I. Sadiek.** has contributed to Formal analysis; Investigation; Visualization; Conceptualization; and Writing – original draft, **A. J. Fleisher** has contributed to Formal analysis; Investigation; Validation; Visualization; and Writing – original draft;, **J. Hayden** has contributed to Investigation; Data curation; Software; Methodology; and Writing – review & editing, **X. Huang** has contributed to Methodology, **A. Hugi** has contributed to Resources, **R. Engeln** has contributed to Writing – review & editing, **N. Lang** has contributed to Project administration; and Writing – review & editing, **J. H. van Helden** has contributed to Resources; Writing – review & editing; Conceptualization; and Supervision.

**Data availability.** Relevant data that supports our experimental findings is available as Supplementary Information and online at INPTDAT repository of the INP (https://www.inptdat.de).

**References**

1. IEA (2021), *Ammonia Technology Roadmap*, IEA, Paris https://www.iea.org/reports/ammonia-technology-roadmap, License: CC BY 4.0.

2. Erisman, J. W., Sutton, M. A., Galloway, J., Klimont, Z. & Winiwarter, W. How a century of ammonia synthesis changed the world. *Nat. Geosci.* **1,** 636–639 (2008).

3. Hong, J., Prawer, S. & Murphy, A.B.. Plasma Catalysis as an alternative route for ammonia production: status, mechanisms, and prospects for progress. *ACS Sustainable Chem. Eng.* **6,** 15–31 (2018).

4. Bogaerts, A. & Neyts, E. C. Plasma technology: an emerging technology for energy storage. *ACS Energy Lett.* **3,** 1013–1027 (2018).

5. Touchard, S., Mougenot, J., Rond, C., Hassouni, K. & Bonnin, X. AMMONX: a kinetic ammonia production scheme for EIRENE implementation. *Nucl. Mater. Energy* **18,** 12–17 (2019).

6. Gordiets, B., Ferreira, C. M., Pinheiro, M. J. & Ricard, A. Self-consistent kinetic model of low-pressure - flowing discharges: II. surface processes and densities of N, H, species. *Plasma Sources Sci. Technol.* **7,** 379–388 (1998).

7. van Helden, J. H., et al. Detailed study of the plasma-activated catalytic generation of ammonia in $N_2$–$H_2$ plasmas. *J. Appl. Phys.* **101,** 043305 (2007).

8. Devasia, D., Das, A., Mohan, V. & Jain, P. K. Control of chemical reaction pathways by light-matter coupling. *Annu. Rev. Phys. Chem.* **72,** 423–443 (2021).

9. Bogaerts, A., et al. The 2020 plasma catalysis roadmap. *J. Phys. D. Appl. Phys.* **53,** 443001 (2020).

10. Vankan, P., Rutten, T., Mazouffre, S., Schram, D. C. & Engeln, R. Absolute density measurements of ammonia produced via plasma-activated catalysis. *Appl. Phys. Lett.* **81,** 418–420 (2002).

11. Puth, A., et al. Spectroscopic investigations of plasma nitrocarburizing processes using an active screen made of carbon in a model reactor. *Plasma Sources Sci. Technol.* **27,** 075017 (2018).

12. Dalke, A., et al. Solid carbon active screen plasma nitrocarburizing of AISI 316L stainless steel: influence of $N_2$–$H_2$ gas composition on structure and properties of expanded austenite. *Surf. Coat. Technol.* **357,** 1060–1068 (2019).

13. Hempel, F., Davies, P. B., Loffhagen, D., Mechold, L. & Röpcke, J. Diagnostic studies of $H_2$–Ar–$N_2$ microwave plasmas containing methane or methanol using tunable infrared diode laser absorption spectroscopy. *Plasma Sources Sci. Technol.* **12,** S98–S110 (2003).

14. Phillips, M. C., Myers, T. L., Johnson, T. J. & Weise, D. R. *In-situ* measurements of pyrolysis and combustion gases from biomass burning using swept wavelength external cavity quantum cascade lasers. *Opt. Express* **28,** 8680–8700 (2020).

15. Witsch, D., et al. The rotationally resolved infrared spectrum of TiO and its isotopologues. *J. Mol. Spectrosc.* **377,** 111439 (2021).

16. Golkowski, M., *et al.* Hydrogen-Peroxide-Enhanced Nonthermal Plasma Effluent for Biomedical Applications. *IEEE Transactions on Plasma Science* **40**, 1984-1991 (2012).

17. Abbas, M. A., Pan, Q., Mandon, J., Cristescu, S. M., Harren, F. J. M., & Khodabakhsh, A. Time-resolved mid-infrared dual-comb spectroscopy. *Sci. Rep.* **9**, 17247 (2019).

18. Bergevin, J., *et al.* Dual-comb spectroscopy of laser-induced plasmas. *Nat. Commun.* **9**, 1273 (2018).

19. Villares, G., Hugi, A., Blaser, S. & Faist, J. Dual-comb spectroscopy based on quantum-cascade-laser frequency combs. *Nat. Commun.* **5,** 5192 (2014).

20. Agner, J. A., et al. High-resolution spectroscopic measurements of cold samples in supersonic beams using a QCL dual-comb spectrometer. *Mol. Phys.* **120,** e2094297 (2022).

21. Komagata, K. N., Wittwer, V. J., Südmeyer, T., Emmenegger, L. & Gianella, M. Absolute frequency referencing for swept dual-comb spectroscopy with midinfrared quantum cascade lasers. *Phys. Rev. Res.* **5,** 013047 (2023).

22. Klocke, J. L., et al. Single-shot sub-microsecond mid-infrared spectroscopy on protein reactions with quantum cascade laser frequency combs. *Anal. Chem.* **90,** 10494–10500 (2018).

23. Lepère M., et al. A mid-infrared dual-comb spectrometer in step-sweep mode for high-resolution molecular spectroscopy. *J. Quant. Spectrosc. Radiat. Transf.* **287,** 108239 (2022).

24. Yurchenko, S. N., Barber, R. J. & Tennyson, J. A variationally computed line list for hot $NH_3$. *Mon. Notices Royal Astron. Soc.* **413,** 1828–1834 (2011).

25. Gordon, I. E., et al. The HITRAN2020 molecular spectroscopic database. *J. Quant. Spectrosc. Radiat. Transf.* **277,** 107949 (2022).

26. Hong, J. et al. Corrigendum: Kinetic modelling of $NH_3$ production in $N_2$–$H_2$ non-equilibrium atmospheric-pressure plasma catalysis (2017 J. Phys. D: Appl. Phys. 50 154005). *J. Phys. D Appl. Phys.* **51,** 109501 (2018).

27. Ben Yaala, M., et al. Plasma-activated catalytic formation of ammonia from $N_2$–$H_2$: influence of temperature and noble gas addition. *Nucl. Fusion* **60,** 016026 (2020).

28. Akima, H. A new method of interpolation and smooth curve fitting based on local procedures. *J. Assoc. Comput. Machin.* **17,** 589–602 (1970).

29. Vankan, P., Schram, D. C. & Engeln, R. Relaxation behavior of rovibrationally excited $H_2$ in a rarefied expansion. *J. Chem. Phys.* **121,** 9876–9884 (2004).

30. Itikawa, Y. Cross sections for electron collisions with ammonia. *J. Phys. Chem. Ref. Data* **46,** (2017).

31. Rowlinson, J. S. The second virial coefficients of polar gases. *Trans. Faraday Soc.* **45,** 974–984 (1949).

32. Hovis, F. E. & Moore, C. B. Vibrational relaxation of $NH_3(\nu_2)$. *J. Chem. Phys.* **69,** 4947–4950 (1978).

33. Lambert, J. D. Vibration-translation and vibration-rotation energy transfer in polyatomic molecules. *J. Chem. Soc. Faraday Trans. 2 Mol. Chem. Phys.* **68,** 364–373 (1972).

34. Dubé, P. & Reid, J. Vibrational relaxation of the $2\nu_2$ level of $NH_3$. *J. Chem. Phys.* **90,** 2892–2899 (1989).

35. Shultz, M. J. & Wei, J. Infrared, resonance enhanced multiphoton ionization double resonance detection of energy transfer in $NH_3$. *J. Chem. Phys.* **92,** 5951–5958 (1990).

36. Abel, B., Coy, S. L., Klaassen, J. J. & Steinfeild, J. I. State-tostate rotational energy-transfer measurments in the $\nu_2$ = 1 state of ammonia by infrared-infrared double resonance. *J. Chem. Phys.* **96,** 8236–8250 (1992).

37. Klarenaar, B. L. M., et al. Time evolution of vibrational temperatures in a $CO_2$ glow discharge measured with infrared absorption spectroscopy. *Plasma Sources Sci. Technol.* **26,** 115008 (2017).

38. Hannemann M., et al. Langmuir probe and optical diagnostics of active screen N2–H2 plasma nitriding processes with admixture of CH4. *Surface and Coatings Technology* **235**, 561-569 (2013).

39. Sterczewski, L. A., et al. Mid-infrared dual-comb spectroscopy with interband cascade lasers. *Opt. Lett.* **44,** 2113–2116 (2019).

40. Sterczewski, L. A., et al. Terahertz spectroscopy of gas mixtures with dual quantum cascade laser frequency combs. *ACS Photonics* **7,** 1082–1087 (2020).

41. Van Gasse, K., et al. An on-chip III-V-semiconductor-on-silicon laser frequency comb for gas-phase molecular spectroscopy in real-time. Preprint at https://doi.org/10.48550/arXiv.2006.15113 (2020).

42. Vervloessem, E., et al. $NH_3$ and $HNO_x$ formation and loss in nitrogen fixation from air with water vapor by nonequilibrium plasma. *ACS Sustainable Chem. Eng.* **11,** 4289–4298 (2023).

43. Coddington, I., Newbury, N. & Swann, W., *Optica* **3,** 414–426 (2016).

44. Šimečková, M., Jacquemart, D., Rothman, L. S., Gamache, R. R. & Goldman, A. Einstein A-coefficients and statistical weights for molecular absorption transitions in the HITRAN database. *J. Quant. Spectrosc. Radiat. Transf.* **98,** 130–155 (2006).

45. Herzberg, G. *Infrared and Raman Spectra of Polyatomic Molecules* (Lancaster Press, Lancaster, PA, 1945).

46. Gamache, R. R., et al. Total internal partition sums for 166 isotopologues of 51 molecules important in planetary atmospheres: application to HITRAN2016 and beyond. *J. Quant. Spectrosc. Radiat. Transf.* **203,** 70–87 (2017).

47. Al Derzi, A. R., Furtenbacher, T., Tennyson, J., Yurchenko, S. N. & Császár, A. G. MARVEL analysis of the measured high-resolution spectra of $^{14}NH_3$. *J. Quant. Spectrosc. Radiat. Transf.* **161,** 117–130 (2015).

48. Coles, P. A., Yurchenko, S. N. & Tennyson, J. ExoMol molecular line lists – XXXV. A rotation-vibration line list for hot ammonia. *Mon. Notices Royal Astron. Soc.* **490,** 4638–4647 (2019).

49. Polyansky, O. L., et al. Calculation of rotation-vibration energy levels of the ammonia molecule based on an *ab initio* potential energy surface. *J. Mol. Spectrosc.* **327,** 21–30 (2016).

50. Philip, R. & Bunker, P. J. *Molecular Symmetry and Spectroscopy* (NRC Research Press, Ottawa, 2006).

51. Possolo, A. & Iyer, H. K. Invited article: concepts and tools for the evolution of measurment uncertianty. *Rev. Sci. Instrum.* **88,** 011301 (2017).

52. Aroui, H., Nouri, S. & Bouanich, J.-P. $NH_3$ self-broadening coefficients in the $\nu_2$ and $\nu_4$ bands and line intensities in the $\nu_2$ band. *J. Mol. Spectrosc.* **220,** 248–258 (2003).

53. Down, M. J., et al. Re-analysis of ammonia spectra updating the HITRAN $^{14}NH_3$ database. *J. Quant. Spectrosc. Radiat. Transf.* **130,** 260–272 (2013).